# Challenging packaging limits and infectivity of phage λ

Elmar Nurmemmedov[1,a], Martin Castelnovo[2], Elizabeth Medina[3], Carlos Enrique Catalano[3], and Alex Evilevitch[1,4*]

1) *Department of Biochemistry, Center for Chemistry and Chemical Engineering, Lund University, Box 124, SE-221 00, Lund, Sweden*

2) *Laboratoire de Physique, Ecole Normale Superieure de Lyon, 46 Allée d'Italie, 69364 Lyon Cedex 07, France*

3) *Department of Medicinal Chemistry, University of Washington School of Pharmacy, H172 Health Sciences Building, Box 357610, Seattle, WA 98195, USA*

4) *Department of Physics, Carnegie Mellon University, 5000 Forbes Ave, Pittsburgh, PA 15213, USA*

\* To whom correspondence should be addressed

e-mail: alexe@andrew.cmu.edu

[a] Present address: Division of Hematology/Oncology, Children's Hospital Boston, Harvard Medical School, Boston, MA 02115, USA.

**ABSTRACT**

The terminase motors of bacteriophages have been shown to be among the strongest active machines in the biomolecular world, being able to package several tens of kilobase pairs of viral genome into a capsid within minutes. Yet these motors are hindered at the end of the packaging process by the progressive build-up of a force resisting packaging associated with already packaged DNA. In this experimental work, we raise the issue of what sets the upper limit on the length of the genome that can be packaged by the terminase motor of phage λ and still yield infectious virions, and the conditions under which this can be efficiently performed. Using a packaging strategy developed in our laboratory of building phage λ from scratch, together with plaque assay monitoring, we have been able to show that the terminase motor of phage λ is able to produce infectious particles with up to 110% of the wild-type (WT) λ-DNA length. However, the phage production rate, and thus the infectivity, decreased exponentially with increasing DNA length, and was a factor of $10^3$ lower for the 110% λ-DNA phage. Interestingly, our *in vitro* strategy was still efficient in fully packaging phages with DNA lengths as high as 114% of the WT length, but these viruses were unable to infect bacterial cells efficiently. Further, we demonstrated that the phage production rate is modulated by the presence of multivalent ionic species. The biological consequences of these finding are discussed.

**INTRODUCTION**

Viruses are among the simplest biological organisms. They typically consist of a viral genome within a protein container, called a capsid, whose function is to protect the genome and to provide a specific strategy for the early steps of infection. Despite their simplicity, viruses exhibit extreme diversity and are highly prone to mutations. This limits the efficiency of current anti-viral therapies, which are often focused on the specificity of viral replication. The past decade has seen the emergence of a new interdisciplinary approach, called "Physical Virology", providing tremendous opportunities for the elucidation of the general physical mechanisms involved in virus development and infection. For example, in double-stranded (ds) DNA viruses, the genome is often hundreds of times longer than the dimensions of the capsid into which it is packaged, and this results in high internal pressure due to the repulsive negative charges on the densely packaged DNA. Indeed, high mechanical pressure in dsDNA bacterial viruses (phages) has recently been demonstrated (1, 2). During infection, the phage binds to the receptor on the bacterial cell surface and the pressure injects the genome into the cell, at least in part, while the empty capsid remains outside the cell. This initiates a series of events, eventually leading to hundreds of viral genomes, pre-assembled capsids and viral tails that are the direct precursors to the assembled, infectious virus. During phage assembly, a motor complex (the terminase enzyme), specifically recognizes viral DNA then binds to the portal complex situated at a unique vertex of the icosahedral capsid; the portal is a ring-like structure that provides a hole through which DNA enters the capsid during assembly and exits during infection (3-6). The terminase motors, which utilize ATP hydrolysis to drive DNA into the capsid, are among the most powerful biological motors characterized to date (3-6). For instance, the bacteriophage λ terminase motor packages DNA to near-crystalline density, which generates an internal pressure in excess of 20 atmospheres. Commensurate with this requirement, the λ terminase motor exerts forces greater than 50 piconewtons during the terminal stages of genome packaging (2, 4, 7-9). Similar features are observed in many bacteriophages and the eukaryotic herpes virus groups. In the latter case, genome packaging occurs in the nucleus of an infected cell, following a pathway that is remarkably similar to that of the phage system, i.e., a terminase motor specifically packages viral DNA into the interior of a pre-assembled procapsid through a unique portal vertex (10-12).

Using phage λ, Evilevitch and co-workers provided the first experimental *in vitro* evidence that internal pressure, often greater than 50 atm, ejects viral DNA from the capsid into the host cell (1, 13). Since the internal capsid pressure is mainly the result of electrostatic repulsion between neighboring DNA strands, we found that the pressure is markedly reduced in the presence of multivalent cations (due to the screening of negative charges on the DNA), since capsids are permeable to water and salts (14). Similar results have been observed in single-molecule laser tweezer experiments (4). Consistent with this, we also found that internal pressure increases strongly with increasing length of the DNA packaged into the capsid, as the negative charge density increases (13). This may explain the observation *in vivo* that a λ genome can only be 8.8 % longer than the wild-type λ genome and still yield an infectious virus *in vivo* (15).

DNA packaging is used for molecular cloning in so-called cell extract "packaging kits", where non-viral genes are inserted into the phage chromosome, packaged *in vitro*, delivered and expressed in bacterial cells (16). This technique has played a revolutionary role in the development of molecular biology and nanotechnology (17). However, while powerful, the size of an inserted gene is limited by the force of the terminase motor and its capacity to package oversized genomes into the capsid interior. During the early steps of the packaging process, the repulsive force exerted by the packaged DNA is smaller than the packaging force generated by the terminase motor. However, as the length of packaged DNA increases, repulsive forces increase and eventually exceed the maximum force that the motor can generate. Thus, the efficiency of viral packaging and virion synthesis are strongly dependent on the DNA length, ionic strength conditions, and types of cations included in the reaction mixture. $Mg^{2+}$ and spermidine$^{3+}$ can stabilize condensed DNA by reducing DNA–DNA repulsion. This may influence the rate of packaging and thus the number of successful packaging events. To our knowledge, no systematic or quantitative *in vitro* measurements have been made to explore the dependence of viral infectivity on the length of the packaged genome and the salt concentration, and the consequences for both controlled laboratory viral assembly (e.g. genetic engineering) and *in vivo* virus assembly. Indeed, this is not possible in the commercially available packaging kits that utilize crude cell extracts.

Catalano and co-workers recently developed an *in vitro* virus assembly system in which an infectious virus was assembled from purified components in a rigorously defined biochemical reaction mixture (5). This allows detailed and quantitative interrogation of each step of the assembly pathway and manipulation of the reaction conditions in a strictly controlled manner. In this work, we exploit this system to investigate the correlation between the internal genome pressure, governed by both the genome length and ionic conditions, and virus infectivity. Our goal was to determine the maximum length of the packaged DNA in phage λ by packaging oversized DNA, relative to the wild-type (WT) λ genome length of 48,502 bp, under various ionic strength conditions. The length of DNA was systematically varied by incorporation of non-specific plasmid DNA into the WT λ genome, and the packaging rate and efficiency were determined by several *in vitro* and *in vivo* assays. The results of these studies are discussed with respect to viral genome packaging *in vivo* and molecular cloning *in vitro*.

**RESULTS**

We constructed three oversized phage λ chromosomes: 51,206 bp (105%), 53,892 bp (110%) and 55,340 bp (114%) long, relative to the WT λ-genome length of 48,520 bp (corresponding to 100%), as described in Materials and Methods section. Briefly, we inserted linearized bacterial plasmids of defined lengths into the *Xba*I-digested λ chromosome (*Xba*I is a unique restriction site located at bp 24,511 of the λ sequence), without disturbing essential λ genes (Figure 1). The inserts contain two markers that allow detection of viral DNA entry into the host cell; lacZ for blue-clear bacterial colony detection, and ampR for ampicillin resistance (Figure 2). The oversized λ genomes were used as packaging substrates in our defined *in vitro* virus assembly system, as described previously (5). The packaging buffer contained 20 mM Tris-HCl, pH 7.5, 4.5 mM $MgCl_2$ and 2 mM $spermidine^{3+}$ (buffer 0), which is optimal for the assembly of an infectious virus using a WT (100%) λ genome (5). As anticipated, infection of *E. coli* JM83 cells ($ampR^-$, $lacZ^-$) with the *in vitro* assembled WT virus did not afford any bacterial colonies when plated on ampicillin-containing plates, because the WT genome does not contain the ampR gene. We next examined particle assembly and the infectivity of viruses assembled with each of the oversized genomes. As before, the genomes were used as packaging substrates in our *in vitro* virus assembly system, and *E. coli* JM83 was infected with the *in vitro* assembled virus. Successful packaging of the oversized genome into the procapsid, addition of a viral tail, and subsequent injection of the viral DNA into the host cell by the infectious assembled virus was indicated by the presence of blue colonies (lacZ) that grow in the presence of ampicillin (ampR) (Figure 2). The data presented in Table 1 demonstrate that oversized genomes up to 114% of the WT genome length can be successfully packaged *in vitro*, and that the virus is able to inject the oversized DNA into the host cell to render them ampicillin-resistant. Importantly, control experiments, in which ATP was omitted from the virus assembly reaction, and thus cannot package DNA, did not yield any colonies on ampicillin-containing plates, indicating that virus particle assembly is required for DNA entry into the cell. Finally, we investigated whether salt concentration affected *in vitro* virus assembly or infectivity, as follows. Each of the genomes was used as packaging substrate, as described above, except that spermidine was omitted from the reaction (buffer 1), or the concentration of $Mg^{2+}$ was reduced to 2 mM (buffer 2). (The terminase packaging motor has a strict divalent metal requirement, which precluded the complete omission of $Mg^{2+}$ from the reaction mixture (5)). All the oversized genomes could be packaged into infectious viruses *in vitro* under all buffer conditions examined (see below).

While the above results demonstrate that oversized genomes can be assembled into infectious viruses *in vitro*, they do not address the question of whether the oversized genomes can be replicated and packaged into infectious viruses *in vivo*. To directly address this question, we turned to a standard viral plaque assay. Each of the genomes was used to assemble an infectious virus, as described above. The *in vitro* assembled virus was then used to infect *E. coli* LE392 cells, and the number of infectious virus particles quantified by plating on a bacterial lawn (18). Each of the observed plaques results from the infection of a single cell with a virus particle (18), and provides a direct method of quantifying the number of infectious virus particles in the assembly mixture. Further, the injected genome must be competent for multiple rounds of replication and infectious virus particle assembly *within the host cell* for a plaque to be visible. Thus, the generation of a viral plaque requires that the genome is competent for the assembly of an infectious virus both *in vitro* and *in vivo*.

The data presented in Table 1 demonstrate that the 105% and 110% oversized genomes provide a functional substrate for DNA replication and virus development *in vivo*, as evidenced by the appearance of viral plaques on the bacterial lawn. In contrast, the 114% genome substrate failed to yield visible plaques, despite the fact that the elongated genome can be packaged and subsequently delivered to the cell *in vitro* (Table 1). This indicates that a 114% long λ genome can be packaged under defined reaction conditions *in vitro*, but that the cellular cytoplasm does not provide an environment conducive to the assembly of an infectious virus with this genome, at least over the multiple infection cycles required to visualize a plaque. We have thus identified the limit of λ genome length that can be efficiently packaged *in vivo*.

While the number of plaques reflects the efficiency of virus assembly *in vitro*, the size of the viral plaque is related to the rate of virus particle assembly *in vivo* – the faster (more efficiently) the virus replicates *in vivo*, the larger the plaque. Figure 3 shows a direct comparison between the 100% and 110% λ genome phages plated on the same LE392 plate, which clearly demonstrates a significant difference in plaque size. As all other factors remain unchanged, and since genome packaging is the rate-limiting step in the concerted viral assembly process (19), we interpret the dramatic reduction in viral replication rate as a reflection of a decrease in the rate of genome packaging *in vivo*. This observation is however not trivial because there are more factors contributing to the plaque size at a given time.

The above interpretation assumes that the input genome was faithfully replicated and packaged *in vivo* to yield infectious, over-packaged virions. We considered the possibility that the oversized genomes were successfully packaged *in vitro* and injected into the host cell, but the duplex length was somehow shortened during replication *in vivo* to generate a viable virus. To directly examine this possibility, the DNA was extracted from the plaques derived from packaging the 105% and 110% oversized genomes. The size of the virion DNA was confirmed with pulsed-field gel electrophoresis, and Figure 4 shows that the input genome is faithfully replicated and packaged into infectious virus particles *in vivo*.

We next quantified the efficiency of virus assembly *in vitro* as a function of genome length. For these studies we utilized our standard reaction buffer 0 (20 mM Tris-HCl pH 7.5, 2 mM spermidine and 4.5 mM $MgCl_2$), and the assembly reaction was allowed to proceed for 2 hours to ensure that it had reached completion. Virus assembly was quantified by plaque assay. The plates were incubated for 12 hours at 37°C, sufficient for the plaques to appear. As can be seen in Figure 5 (red dashed line), the number of plaques, and thus the efficiency of packaging, shows an exponential decrease with increasing packaged λ-DNA length. The number of assembled virions decreased exponentially by a factor of 100 for 5% extra packaged DNA length, relative to the WT. The titer fell by 1000 times when the λ-DNA length was increased from 100% to 110%, and would presumably be 10,000 times lower for the 114% λ-DNA sample. The latter would reduce the transfection efficiency to less than that required for detection by the plaque assay. Similar relative drop in the packaging efficiency was also found for in vivo packaging in ref. (15), when increasing λ-DNA length from 105% to 108.8%. However, the initial drop in the efficiency when going from 100% to 105% λ-DNA length was smaller than in the present study. This is explained by the fact that salt conditions cannot be controlled in an in vivo packaging process unlike in our in vitro packaging study. As we have shown here, packaging efficiency will have a strong ionic dependence for smaller variation in the packaged DNA length around 100% WT λ-DNA length (see also discussion below).

Finally, we quantified the effect of salt concentration on the efficiency of virus assembly *in vitro*. Specifically, we decreased the concentrations of spermidine$^{3+}$ (Figure 5, green dashed line) and Mg$^{2+}$ (Figure 5, blue dashed line) independently, to ensure that any observed effect was the result of a counterion-induced decrease in DNA–DNA repulsion, rather than the effect of a specific salt on the viral protein assembly during packaging. In both cases, decreasing the counterion concentration significantly decreased the yield of infectious viruses *in vitro*, with spermidine exhibiting slightly stronger effects (Figure 5).

**DISCUSSION**

We have examined the assembly of an infectious λ virus *in vitro* under rigorously defined conditions, and have demonstrated that both the rate and extent of particle assembly are strongly affected by the length of the genome substrate. We have further demonstrated that while very large DNA duplexes (114% of WT genome length) can be packaged into a particle and efficiently delivered to a host cell *in vitro*, there is a more stringent limit on virus development *in vivo*. What limits the efficiency of packaging over-sized genomes?

*Effect of DNA length on the packaging efficiency and viral infectivity.* There is an exponential decrease in the yield of infectious viral particles assembled *in vitro* as a function of genome length. The only variable in the assembly reaction is the length of the DNA used as a packaging substrate, and we suggest two reasonable explanations of this finding. First, it might reflect differences in the rate of assembly of the packaging motor complex. The reactions are initiated by terminase, which must bind to the DNA, and the binary complex must then bind to the procapsid to initiate packaging. It is unlikely that a change in the genome length would affect the activation of the packaging motor once assembled. We consider, however, that the rate-limiting step in packaging initiation is related to the free diffusion of the DNA coil in the solution in order to first find a terminase and a free procapsid. From classical polymer physics it can be estimated that the change in diffusion coefficient for two different DNA lengths, $L_{WT}$ and $L$, scales as $D(L) \approx D(L_{WT})(L_{WT}/L)^\nu$, where the exponent of polymer statistics is $\nu = 1/2$ or $3/5$ for an ideal Gaussian chain or a swollen chain (20). As a consequence, we anticipate a reduction of only 5% in the diffusion coefficient of the DNA coil when its length is increased from the WT length to 110% of the WT length. This modest change in diffusion coefficient is unlikely to be responsible for the large decrease in virus assembly (of the order of $10^3$) observed in our experiments.

An alternative explanation is that the observed decrease in the rate and extent of infectious particle assembly reflect a significant increase in the time required to package the oversized duplex, and a decrease in the efficiency of particle completion in the face of an over-packaged capsid, respectively (Figure 1). As we argue below, this exponential decay is related to the dynamic properties of the terminase packaging motor under an external load (*i.e.* the external load does not arise from the motor itself).

The mechanical properties of viral packaging motors have been characterized experimentally at the single molecule level during the past decade (2). Using laser tweezer techniques, it has been shown for phage λ (4) and for phage ϕ29 (21), that if an external load challenges the active packaging process of the terminase motor, the leading order relation between the packaging rate $v$ and this external load $F$ scales as $v \approx v_0 \exp\left(-\frac{F\delta}{kT}\right)$, where $v_0$ is the measured velocity in the absence of the load, and $\delta$ is the step size of the motor, $k$ is Boltzmann's constant and $T$ is the temperature. Note that this relation has been established experimentally as a dynamic intrinsic property of the motor, regardless of DNA filling inside the capsid. The origin of this exponential behavior is intimately related to the decrease in power stroke efficiency of the terminase motor under an opposing external load (22) and the increased frequency of slips of the DNA as it is being packaged at higher loading forces (2). The equations require an irreversible step, which is the case here because cos-cleavage of λ-DNA is irreversible.

In the absence of an external load, it has also been observed that packaging rate slows down dramatically after the first 30% of the WT λ-genome has been packaged. This is the signature of the internal force build-up that resists active packaging by the motor, arising from the DNA confinement in the capsid. Slowing of the packaging rate is qualitatively described by a similar relation between velocity and force $v(L) \approx v_0 \exp\left(-\frac{F_{int}(L)\delta}{kT}\right)$, where $L$ is the packaged length and $F_{int}(L)$ is the internal force associated with the confinement of DNA of length $L$ within the capsid. Therefore, we used this relation to estimate the average time, $T(L)$, required to package DNA of length $L$, $T(L) = \frac{\int_1^L dx \exp\left(\frac{F_{int}(x)\delta}{kT}\right)}{v_0}$.

For DNA around the same length as that of WT (48.5kb) or larger, Fuller *et al.* (4) showed that the internal force is large (25 pN at 90% of WT DNA length). Assuming, for the sake of simplicity, that the relation between the internal force and the genome length is linear above the WT DNA length in the range addressed in our experiments, the average time required to package a length $L$, that is larger than the WT length $L_{WT}$, scales roughly as suggested both by the experimental results of Fuller *et al.* (4) and by DNA packaging models, such as the inverse spool model (23-26):

$$T(L) \cong T(L_{WT}) + \frac{kT \exp\left(\frac{A(L-L_{WT})\delta}{kT}\right)}{Av_0\delta} \quad \text{(Eq. 1)}$$

where $A$ is the coefficient of the linear relationship between the internal force and the genome length. The dynamic properties of the terminase motor therefore predict an exponential increase in the time required to fully package oversized genomes. As a consequence, the rate of virus assembly $J(L) = 1/T(L)$ will decay exponentially with the genome length. This conclusion implicitly assumes that the rate-limiting step for the production of $N$ viruses is the assembly rate, such that $N(L) \approx J(L)T_{exp}$, where $T_{exp}$ is the duration of the packaging experiment. Provided that the incubation time for *in vitro* phage packaging and assembly was 2 hours for all samples, and that the assembly components and enzymes retain their activity during this time period, exponential decay in the production rate leads to the exponential decay of the number of assembled, infectious phage particles, in agreement with the experimental results presented here. Infinitely long packaging times may not even be feasible *in vivo* as cells may run out of ATP required for packaging, and/or the packaging enzymes and viral DNA may be degraded. Indeed, the decrease in plaque size as a function of genome size indicates that the rate of particle assembly is also dramatically slower *in vivo*. Thus, a change in the packaged DNA length of even a few percent, accompanied by an increase in the internal pressure, will have dramatic consequences for viral reproduction and the spread of infection, decreasing by several orders of magnitude.

*Effect of salt on DNA packaging.* The data presented here provide further information on the influence of ionic species on the efficiency of virus assembly under rigorously defined conditions. Our results are consistent with recent single-molecule laser tweezer studies showing that polyamines and $Mg^{2+}$ decrease the internal force resisting packaging (7). Moreover, the single-molecule packaging rate is systematically faster in the presence of multivalent ions, at least when packaging a unit-length WT genome. The consequence of this observation is that the time required to package DNA up to WT genome length should be shorter in the presence of multivalent ions. Our data indicate that the time required to package DNA beyond its WT length, up to 110% of the natural genome, depends strongly on a combination of the dynamic properties of the motor and the force resisting packaging. Examination of Fuller's measurements (7) suggests that the relation between the internal force and packaged DNA length around the WT length is still linear and is only slightly dependent on the ionic conditions. Therefore, the previous scaling estimation of the time required for packaging of oversized λ-DNA can be reproduced.

These two observations lead to the following conclusion: the initial packaging of DNA up to WT genome length is faster in the presence of multivalent ions, but the duration of the later stages of packaging beyond the WT genome length should still increase exponentially with genome length. As a consequence, our approach predicts that the addition of multivalent ions will increase the phage production rate at constant genome length, while maintaining approximately the same exponential decay with increasing genome length. This matches our experimental data qualitatively (Figure 5) and further confirms our interpretation of the overpackaging data. Thus, the phage production rate is increased with more efficient cation induced reduction of DNA-DNA repulsive force resisting packaging. This is achieved by higher cation charge (as can be seen by comparing spermidine$^{3+}$ and $Mg^{2+}$, both naturally present in the E. coli cytoplasm) or an increase in the overall ionic strength (Figure 5). Ionic strength values, *I*, for all three buffers were 13.5 mM, 18 mM and 25.5 mM, respectively, excluding the contribution from the Tris buffer.

It is interesting to note that with increasing DNA length, the effect of salt on the packaging efficiency becomes weaker. It can be seen in Figure 5 that the addition of 2 mM spermidine$^{3+}$ (with 4.5 mM Mg$^{2+}$ in both buffers 0 and 1) increases the titer by ≈20 times for 100% λ-DNA length, but only by ≈8 times for 110% λ-DNA packaging. In Figure 5, it can be seen that increasing the ionic strength from 13.5 mM (buffer 1) to 18 mM (buffer 2), increases the titer by ≈1.5 times for 100% λ-DNA but there is no titer change for 110% λ-DNA. Thus, the phage titer curves at different ionic conditions shown in Figure 5 are converging with increasing DNA length.[1] This observation is consistent with well-documented observations using osmotic stress experiments (27-29), where it was shown that DNA-DNA interactions in dense hexagonal phases of DNA (similar to DNA densities in overpackaged phage λ) are dominated by the hydration forces and have little dependency on the identity of the salt. Furthermore, high DNA packaging densities lead to several times higher counter-ion concentration at the interior of the capsid, compared to the bulk, making the internal forces less sensitive to the external ion content (30, 31).

---

[1] It should be noted that y-axis, showing pfu/mL, is plotted on a log-scale, making the lines look essentially parallel. However, the differences in phage titer described above between different buffers at various DNA lengths are significant as all 3 exponential curves are asymptotically approaching zero.

**CONCLUSIONS**

Elegant studies by Feiss and co-workers (15) have demonstrated that genomes as long as 1.08 could be packaged from di-lysogens *in vivo*, which sets an upper limit on productive virus development in the cell. The goal of this work was to (i) define the physical limit of DNA that can be inserted into a capsid by the terminase motor and (ii) to elucidate the role of the internal genome pressure on virion synthesis and viral infectivity. These studies take advantage of biochemically defined *in vitro* virus assembly assay recently developed in the Catalano lab. The genome pressure was varied by changing two parameters: DNA length and salt condition in the virus packaging-assembly buffer. The key to these measurements was our recently developed system for the *in vitro* assembly of infectious phage λ, using individually purified components in a rigorously defined biochemical assay (5). For the first time, we were able to quantitatively measure a dramatic reduction in the phage production rate associated with the packaging of DNA longer than the WT genome in the phage λ capsid. These measurements reveal important biological implications. The first is that it is possible to package DNA up to 114% of the WT genome length for phage λ. Moreover, by using different ionic conditions, we demonstrated that overpackaging is a rather robust feature.

The results of our investigations have implications with respect to the presence and the origin of a *maximum length of DNA that can be packaged within a viral capsid.* Indeed, we found that longer DNA than WT genome length can be packaged. However, the rate of packaging, and thus the number of phages produced in a given time, decreases exponentially with increasing genome length. More precisely, a 1% increase in the length of the packaged genome above WT, leads to ten-fold decrease in the viral titer. Moreover, since salt decreases the internal pressure, adding extra salt increases the titer. With higher packaging density (above 110% of WT λ-DNA) the hydration force dominates the DNA-DNA interaction and the pressure is only weakly affected by the addition of counterions, making the packaging rate essentially insensitive to salt conditions. As a consequence, we found that the maximum length of DNA that could be packaged is not limited by the salt conditions of the virus assembly environment. Instead, it is limited by the maximum force that the terminase motor can exert. Remarkably, we also found that maximum length of DNA that can be packaged *in vivo* is between 110 and 114% of WT λ-DNA, however the packaging efficiency was very low.

These observations illustrate that the dynamics of the viral packaging motor is exponentially dependent on the genome length. The motor packaging rate approaches zero asymptotically as the internal pressure increases. Thus, there is no mechanical limit on the length of DNA that can be packaged (assuming that the capsid can withstand any force). This has interesting implications on DNA packaging by the "headful" phages. These viruses initiate packaging at a specific "pac" site in the concatemer, but package longer than unit length duplexes (102-104% genome length). The mechanism that defines the downstream cut site remains unclear, but slowing of the terminase packaging motor certainly plays a major role. Indeed, a simple model is that upon packaging a headful of DNA, the terminase motor slows to essentially zero as the pressure of the packaged DNA overcomes the power of the motor. This provides time for the slow endonuclease activity of terminase to cut the duplex to complete the packaging process.

In the case of unit-length genome packaging viruses such as $\lambda$, the packaging time becomes infinitely long as DNA length exceeds WT, preventing viruses from assembling within reasonable time (or at all) to complete their replication cycle. This suggests that the length of the packaged genome has been evolutionarily optimized to packaging densities where the effect of salt on the motor efficiency is minimized, making packaging more robust and less prone to changes in the cellular environment. Most importantly, these findings demonstrate that slight changes in the internal pressure induced by variation in the DNA length, will have fatal consequences for virus synthesis and the spread of infectious particles. This was demonstrated by the dramatic reduction in the number (efficiency) and size (rate) of plaques. These observations strengthen the role of internal pressure in the infectivity process, in agreement with several in vivo studies (32, 33). This study also suggests new strategies for interfering with viral infectivity through small changes in the internal genome pressure in the case of motor-packaged DNA viruses.

At the same time, the fact that overpackaging is possible in various environments might be considered as a potential evolutionary advantage: as phage $\lambda$ would therefore have the ability to modulate its genome length, since both its terminase motor characteristics and its host (the bacteria) are capable of packaging a genome larger than the WT length. More precisely, our finding of an effective upper limit for packaging of the oversized 114% WT DNA means that modulation of genome length is possible up to an extra length of 6.8 kb. Considering that the typical average size of a gene is of the order of 1 kb, phage $\lambda$ therefore has the ability to increase its genome by at least one or two genes. This is a remarkable feature with respect to the adaptation of the virus to its environment.

Based on recent results obtained in our studies (34, 35) and those at other labs (4), on the physical properties of phage λ, we speculate that λ WT length is the optimal result of a balance between a large number of genes, in order to function properly, a long genome length in order to be able to efficiently initiate passive ejection of the genome into the host bacteria, and a genome length short enough that efficient virus production is compatible with the time frame of the cell cycle.

**MATERIALS AND METHODS**

*Construction of oversized phage λ chromosomes.* Oversized phage λ chromosomes, 105%, 110% and 114% of the length of the wild type, were constructed by insertion of linearized bacterial plasmids of defined length into the *Xba*I-digested λ chromosome. *Xba*I is a unique restriction site that is located at 24511 base pairs in an inert gene that seemingly does not affect phage replication. 200 μg of λ DNA (dam$^-$ dcm$^-$) was digested with 20 units *Xba*I at 37 ºC during a period of 12 hours. Digested λ DNA was treated with calf intestinal alkaline phosphatase at 37 ºC for 5 hours. The λ DNA was then purified using a phenol-chloroform purification technique. Bacterial plasmid pUC19 (Amp$^R$ and LacZ$^+$, 2686 base pairs) was used for construction of the 105% and 110% λ chromosomes. Firstly, it was made compatible for the ligation with the digested λ DNA. An adaptor was inserted into its unique *Psc*I site. The adaptor (5'-CATGGCTAGCAACCTAGGAAACTAGT-3' and 5'-CATGACTAGTTTCCTAGGTT GCTAGC -3') carries three restriction sites (*Nhe*I, *Avr*II and *Spe*I), all of which produce sticky ends compatible with those of *Xba*I. Thus, *modified-pUC19* was generated. This was consequently used for the construction of *double-pUC19*, which is basically two fused copies of the same plasmid. Modified-pUC19 was digested with *Nhe*I, purified from 0.8% agarose gel and ligated with itself. Ligation products, amplified in *E. coli*, were enzymatically linearized and separated on 0.8% agarose gel electrophoresis. The DNA fragment corresponding to twice the size of pUC19 was extracted, purified and circularized with T4 DNA ligase. Thus double-pUC19 (5372 base pairs) was created. For the construction of the 114% λ chromosome, plasmid pSV-b-galactosidase (Amp$^R$ and LacZ$^+$, 6820 base pairs) was used. This plasmid has a unique *Xba*I site, which is conveniently located in an inert region.

In order to construct the 105%, 110% and 114% λ chromosomes, 50 μg of each of the modified-pUC19, double-pUC19 and pSV-β-galactosidase plasmids were digested separately with 10 units of *Spe*I (for modified-pUC19 and double-pUC19) and *Xba*I (for pSV-β-galactosidase) at 37 ºC over the course of 12 hours. DNA was subsequently separated on 0.8% agarose gel electrophoresis, and fragments corresponding to 2686, 5372 and 6820 base pairs were extracted and purified. Ligation reactions were subsequently performed with *Xba*I-digested and dephosphorylated λ DNA. 0.8 μg of λ DNA was used per 50 μl of ligation reaction, into which 1 unit of T4 DNA ligase was introduced. A molar ratio of 1:8 (1 for λ DNA, 8 for plasmid) was employed. Reactions were performed at 16 °C for 12 hours. Ligation was accompanied by four negative control reactions, in which the reaction components were sequentially removed: 1) no plasmid insert, 2) no T4 ligase, 3) no λ DNA, and 4) no λ DNA and no T4 ligase. The phenol-chloroform-purified total DNA product from ligation reactions was used as packaging substrate.

*Assembly of oversized λ–DNA phages.* Phage λ was built from scratch from its recombinantly produced protein components in a two-step process. In comparison to the initially described method (1), we have slightly modified this process in terms of DNA and packaging buffer. In the packaging step, the following viral components were sequentially mixed in 25 μl reaction in a controlled ionic environment: 8.48 mM ATP, 95 μM gpD, 424 nM gpFI, 2 μM IHF, 208 nM procapsids, 1 μM terminase and DNA. The packaging process was performed under three different sets of ionic conditions:

Buffer 0          20 mM Tris-HCl pH 7.5, 2 mM spermidine and 4.5 mM $MgCl_2$

Buffer 1          20 mM Tris-HCl pH 7.5 and 4.5 mM $MgCl_2$

Buffer 2          20 mM Tris-HCl pH 7.5, 2 mM spermidine and 2 mM $MgCl_2$

The quantities of DNA used for this step were empirically determined (as described below). After incubation for 2 hours at 25 ºC, the packaging process was considered complete.

Prior to the subsequent assembly step, the viral tail segment was constructed by mixing 417 μM gpFII, 413 μM gpW and 320 nM tail in 10 μl reaction, with 20 mM Tris-HCl pH 7.5, 2 mM spermidine, 3 mM $MgCl_2$, 7 mM β-mercaptoethanol and 25 mM potassium glutamate. This mixture was then added to the packaged viral shells from the previous step and incubated for 2 hours at 37 ºC. This step resulted in a fully infective phage. Subsequently, the complete λ phage was used for in vivo assays. Positive controls were similarly prepared using wild-type λ DNA.

In vivo assays using E. coli

In order to test our viral assembly system, we performed phenol-chloroform extraction of DNA from the packaged viral heads (directly after the packaging step). Extracted DNA was separated on 0.8% agarose gel for qualitative determination. However, this approach was successful only for the wild-type phage, as the DNA yield from the overpackaged phages was too low to be detected at this stage. Therefore, we performed two complimentary *in vivo* tests.

Blue-clear screening was performed in order to confirm the integration of the plasmid DNA into the phage λ chromosome. Assembled phages were used to infect *E. coli* JM83 (AmpR$^-$ LacZ$^-$), grown to an $OD_{600}$ of 1.0 in LB medium supplemented with 10 mM $MgSO_4$ and 0.2% (w/v) maltose. Infected cells were spread on LB agar plates containing 100 µg/ml ampicillin, 40 µg/ml X-gal and 0.5 mM IPTG (isopropyl β-D-1-thiogalactopyranoside). Plates were incubated at 37 ºC for 12 hours. *E. coli* JM83 infected with overpackaged phage λ yielded blue colonies, whereas *E. coli* JM83 infected with the wild-type phage λ did not yield any colonies. This *in vivo* test clearly demonstrated that the presence of phage-delivered Amp$^R$ and LacZ genes provides *E. coli* JM83 with ampicillin resistance and blue pigmentation. It is thus an indicator of the successful construction of oversized phage λ chromosomes.

Plaque assay was performed for quantitative determination of packaging efficiency and *in vivo* infectivity of the overpackaged phages. Assembled phages were diluted 10, 100 and 500 times and mixed with 100 µl *E. coli* LE392, which was grown to an $OD_{600}$ of 1.0 in LB medium supplemented with 10 mM $MgSO_4$ and 0.2% (w/v) maltose. Infected cells were spread on LB agar plates. The plates were incubated at 37 ºC for 12 hours. Pulsed-field gel electrophoresis was used to visualize the DNA and assess the difference in size, using CHEF DR II electrophoresis equipment (Bio-Rad). Purified DNA samples were loaded into 1% agarose gel. The gel was run for 26.5 hours in 0.5x TBE buffer using the following conditions: initial switch of 2.9 seconds, final switch of 4.5 seconds and voltage of 6 V/cm. The gel was later stained with SYBR Gold nucleic acid dye and visualized under 302 nm UV. We observed a clear size difference between wild-type, 105% and 110% λ DNA bands.

*DNA saturation experiments:* The amount of DNA used for the assembly of overpackaged λ phages affected the final yield, *i.e.* the number of plaques, as was previously shown (1). Therefore, we had to establish saturation levels for 105% and 110% DNA empirically. Since the ligation product used as packaging substrate was a mixture of various DNA products, including the oversized DNA product, we used weight units (µg) of total DNA used for saturation level experiments. We designed phage assembly experiments (as described above) where various weight units, between 3 and 40 µg, of DNA substrate were used. Packaging buffer 0 was used for this experiment. These phages were subsequently used for plaque assay, as described above. Plaque numbers were counted and used to estimate the DNA load per packaging reaction that would ensure the oversaturation of packaging mixture with the DNA substrate. These saturation levels were later used in further experiments for the determination of the dependence of packaging efficiency on the ionic strength.

**ACKNOWLEDGMENTS:** We would like to thank William Gelbart and Charles Knobler for fruitful discussions and inspiration, Lars-Olof Hedén for tremendous help with the microbiological part of this project, and John Pettersson for technical help. This work was supported by Carl Trygger's Foundation grant to AE and the Swedish Research Council (VR) grant to AE (grant # 2008-726).

**FIGURE LEGENDS:**

**Figure 1:** Over-sized phage λ chromosomes of length 105%, 110% and 114% were built by insertion of linearized bacterial plasmids of defined length into the *Xba*I-digested λ chromosome. All plasmids contained $Amp^R$ and LacZ genes providing ampicillin resistance and blue pigmentation upon infection of *E.coli* JM83. It is an indicator of successful construction and packaging of oversized phage λ chromosomes in vitro. In parallel, efficiency of the in vivo phage assembly after the bacterial cell infection is monitored by plaque assay, revealing the number of assembled and infectious phage particles determined by the number of plaques per plate.

**Figure 2:** Blue-clear screening was performed in order to confirm integration of the plasmid DNA into the phage λ chromosome. Assembled phages were used to infect *E.coli* JM83 ($AmpR^-$ $LacZ^-$). Infected cells were spread on LB-agar plates. *E.coli* JM83 infected with overpackaged phage λ yielded blue colonies, whereas *E.coli* JM83 infected with wild-type phage λ did not yield any colonies. This *in vitro* assembly test clearly demonstrated that presence of phage-delivered $Amp^R$ and LacZ genes provides ampicillin resistance and blue pigmentation to *E.coli* JM83 (Petri dish photo is shown next to the cartoon). Plaque assay was performed as a quantitative test of the virion assembly efficiency *in vivo*. Assembled phages were mixed with *E.coli* LE392. Infected cells were spread on LB-agar plates and after 12 hours incubation at 37 °C plaques could be observed (Petri dish photo is shown next to the cartoon).

**Figure 3:** Plaque assay for WT (100%) and 110% λ-DNA phages on *E. coli* LE392 plate, illustrating large plaques for 100% DNA phage and small plaques for 110% DNA phage.

**Figure 4:** Pulsed-field gel electrophoresis of DNA extracted from plaques carrying oversized λ genomes.

**Figure 5:** Efficiency of packaging monitored by plaque assay. pfu/mL is plotted on a log-scale versus % of packaged DNA length (100% corresponds to the wild-type λ-DNA length).

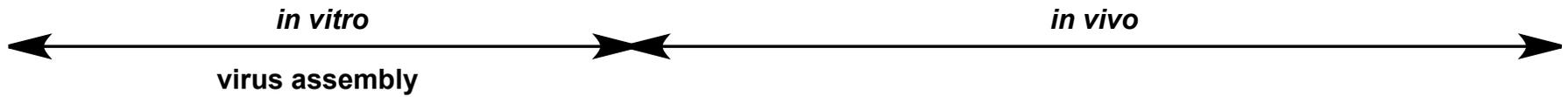

*in vitro* ← virus assembly → *in vivo*

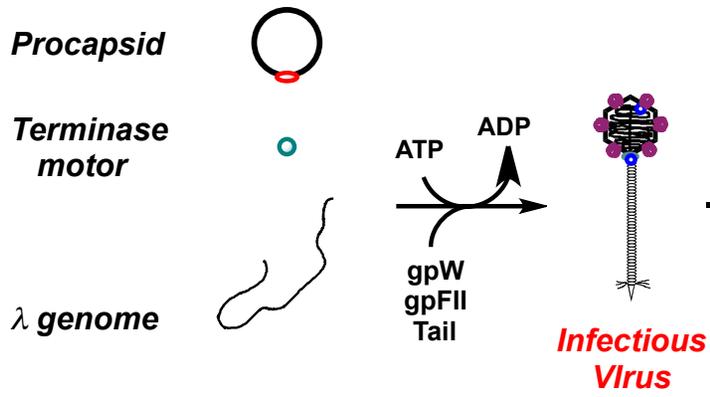

***Procapsid***

***Terminase motor***

***λ genome***

WT = 100% = 48,502 bp
     106% = 51,206 bp
     111% = 53,892 bp
     114% = 55,340 bp

ATP → ADP
gpW
gpFII
Tail

***Infectious Virus***

Infect *E. coli*

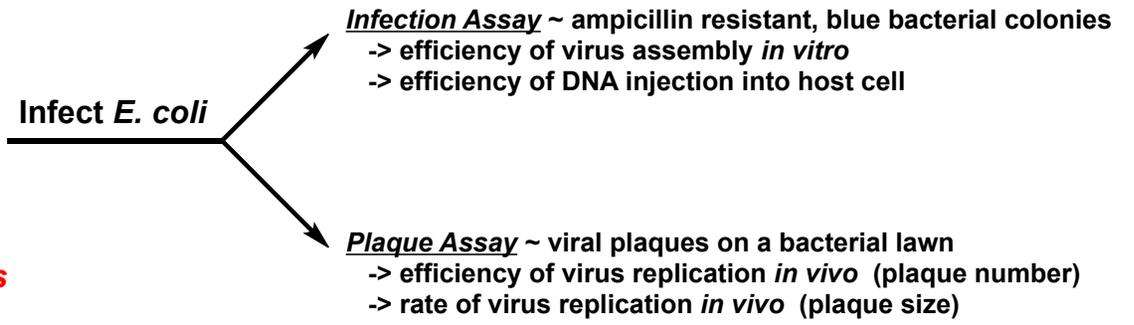

<u>*Infection Assay*</u> ~ ampicillin resistant, blue bacterial colonies
   -> efficiency of virus assembly *in vitro*
   -> efficiency of DNA injection into host cell

<u>*Plaque Assay*</u> ~ viral plaques on a bacterial lawn
   -> efficiency of virus replication *in vivo* (plaque number)
   -> rate of virus replication *in vivo* (plaque size)

*λ genome*

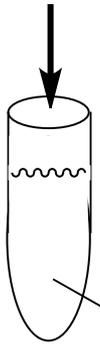

*Virus Assembly Reaction Mix*

→ Virus Particle Assembly →

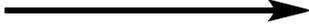

Infect *E. coli*

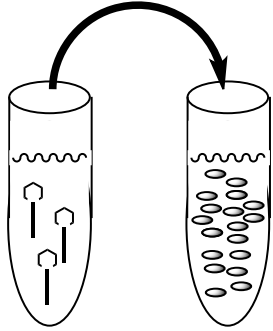

→ Innoculate Bacteria Ampicillin Plates →

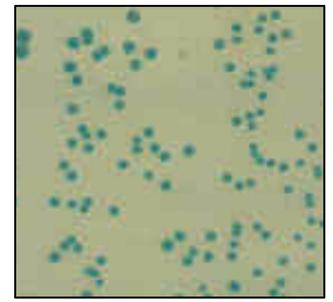

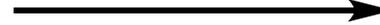

Blue Bacterial Colonies Indicate Particle Assembly and Host Cell Infection *in Vitro*

↓ Plate Bacterial Lawn (plaque assay)

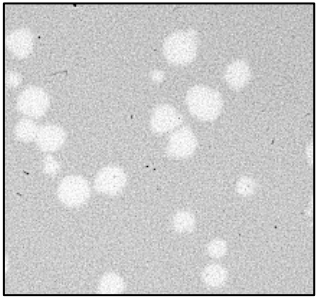

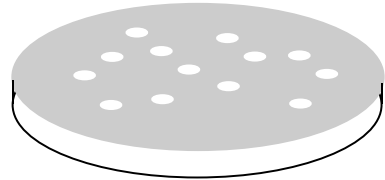

- Plaque Number (pfu) Reflects Assembly Efficiency *in Vitro*

- Plaque Size Reflects Replication Rate *in Vivo*

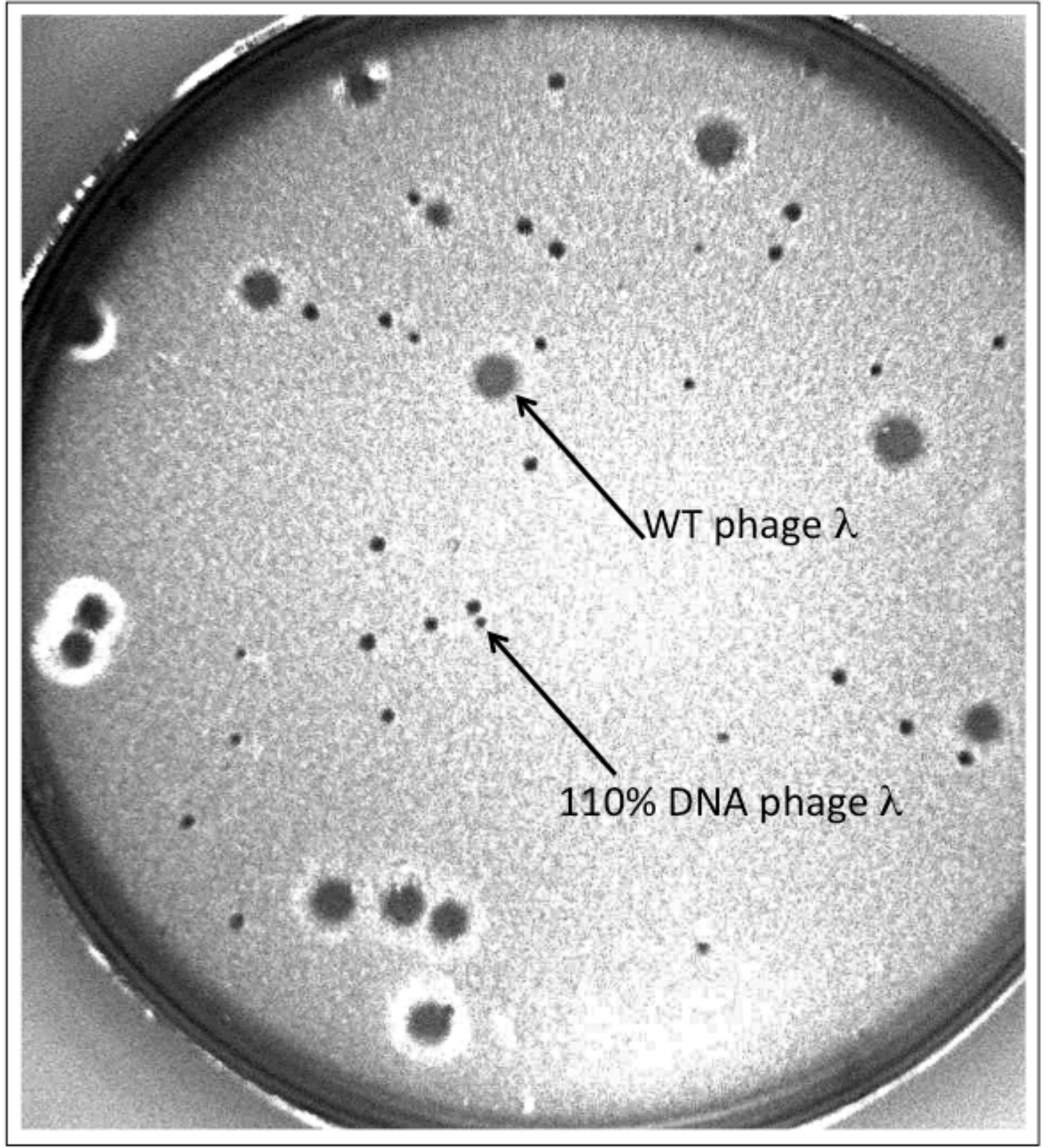

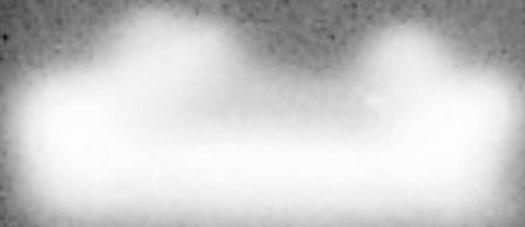
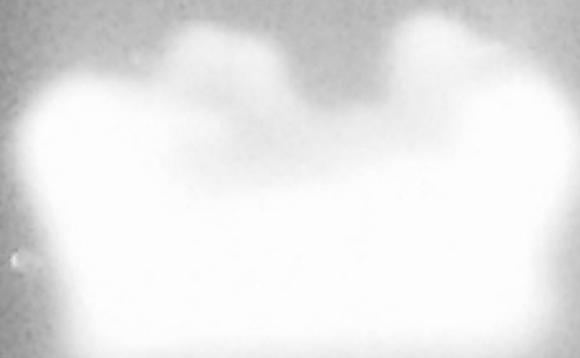
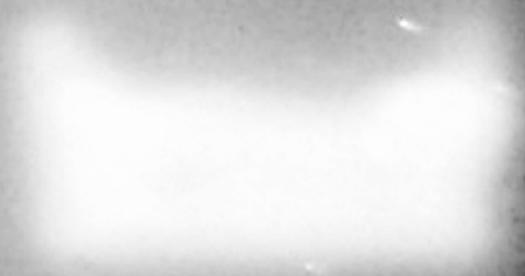

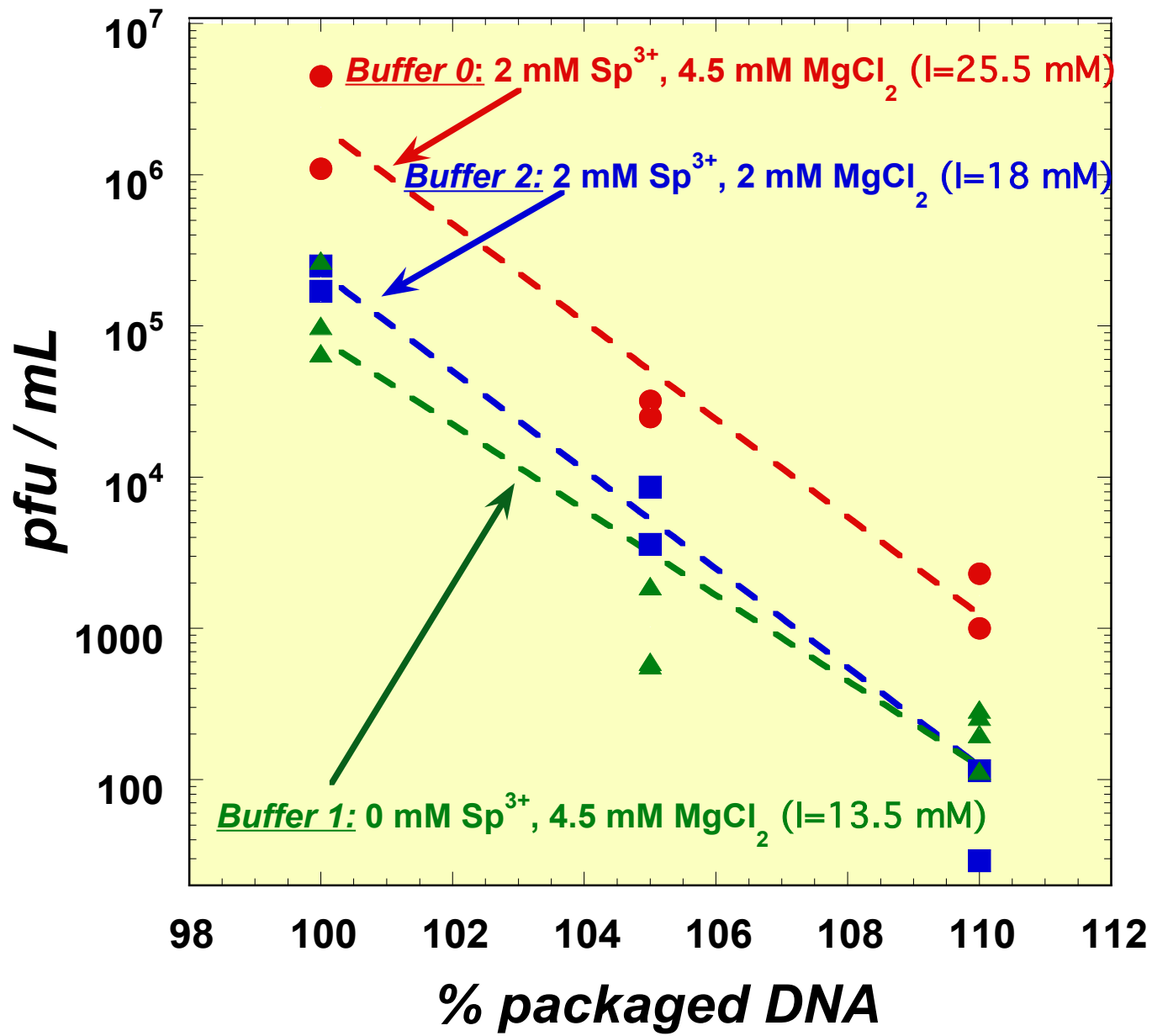